\documentclass[twocolumn,prb,showpacs]{revtex4}
\usepackage{graphicx}
\usepackage{bm}
\usepackage{color}

\begin{document}
\title{Current-current correlations in the three-band model for two-leg CuO ladders}
\author{S.~Nishimoto}
\affiliation{Leibniz-Institut f\"ur Festk\"orper- und Werkstoffforschung Dresden, D-01171 Dresden, Germany}
\author{E.~Jeckelmann}
\affiliation{Institut f\"ur Theoretische Physik, Leibniz Universit\"at Hannover, D-30167 Hannover, Germany}
\author{D.J.~Scalapino}
\affiliation{Department of Physics, University of California, Santa Barbara, California 93106-9530, USA}
\date{\today}
\begin{abstract}
We study current-current correlations in the three-band Hubbard model for two-leg CuO 
ladders using the 
density-matrix renormalization group method. We find that these
correlations decrease exponentially with distance for low doping 
but as a power law for higher doping.  
Their pattern is compatible with the circulating current (CC) phase
which Varma has proposed to explain 
the pseudo-gaped metallic phase in underdoped high-temperature superconductors.
However, for model parameters leading to a realistic ground state in the undoped ladder, the current 
fluctuations decay faster than the d-wave-like pairing correlations in the doped state. Thus 
we conclude that no phase with CC order or dominant CC fluctuations occur in the three-band model 
of two-leg CuO ladders. 
\end{abstract}
\pacs{71.10.Fd, 71.27.+a, 71.10.Hf,74.20.Mn}
\maketitle 

Since the discovery of the high-temperature superconducting copper oxide
compounds, the anomalous behavior of the contiguous ``pseudogap" phase
has been considered a key to understanding the superconductivity mechanism in
these materials. However, the nature of the pseudogap transition and
its order parameter remained a puzzle.  Early $\mu$SR experiments\cite{Sonier01} on
YBa$_2$Cu$_3$O$_{6+x}$ crystals showed evidences for the onset of spontaneous
static magnetic fields near what was called the pseudogap crossover temperature
$T^*(x)$. In addition, different photocurrents for left- and right-circularly
polarized photons in angle-resolved photoemission spectroscopy\cite{Kaminski02}
were reported for Bi$_2$Sr$_2$CaCu$_2$0$_{8+\delta}$. Now recent polarized
neutron scattering\cite{Fauque06} and Kerr effect\cite{Xia08} measurements on
YBa$_2$Cu$_3$O$_{6+x}$ provide new evidence that there is a pseudogap phase
associated with a novel magnetic transition. The neutron scattering
experiments observe a phase characterized by a magnetic order which does not
break translational symmetry and the polar Kerr studies find the phase
transition at $T^*(x)$ which breaks time-reversal symmetry. There
is at present no agreement regarding a theory which encompasses all of these
observations.

Theoretically, ground states, in which circulating currents (CC) 
form spontaneously and thus break the time-reversal symmetry, 
have been found in several models~\cite{Marston02} 
but in the more realistic
$t-J$ model the current-current correlations decrease
exponentially fast on a two-leg ladder.~\cite{Scalapino01} 
However, Varma has argued~\cite{Varma06} 
that the minimal model
for a CC state is a doped three-band Hubbard-type model with one Cu $d$-orbital,
one O $p_x$-orbital, and one O $p_y$-orbital per unit cells. Using a mean-field
approach he has found that a CC ground state is possible in
this model if the Cu-O hopping integral $t_{pd}$ is of the same order of
magnitude as the nearest-neighbor Coulomb interaction $V_{pd}$ between the Cu
and the O orbitals and larger than the energy difference $\Delta_{pd}$
between these orbitals. This CC state, which breaks time-reversal
symmetry but not translational symmetry, is consistent with the neutron
scattering experiments, but further additions\cite{Aji08} to the model are
required to obtain results compatible with the orientation of the moments and
the Kerr rotation results. Moreover, since the interaction between particles is
strong, the mean-field approach cannot reliably determine if a CC phase
really exists in the three-band model.

Several studies of this model have been carried out 
to check Varma's theory
using methods for 
strongly correlated systems. 
Power-law current-current correlations have been observed in CuO chains.~\cite{Srinivasan02} 
A related ``staggered flux'' phase but no CC phase has been found in 
the weak- and strong-coupling phase diagram of undoped two-leg 
ladders.~\cite{Lee05}
Recently, a phase with dominant orbital current fluctuations
has been reported in the weak-coupling phase diagram of doped two-leg
ladders.~\cite{Chudzinski07} 
However, exact diagonalizations of small square clusters~\cite{Greiter07} 
show no evidence for CC patterns in the ground state.
Thus, the existence of a CC order or dominant CC fluctuations in the three-band model
is still an open question. 

\begin{figure}[b]
\includegraphics[width= 6.5cm,clip]{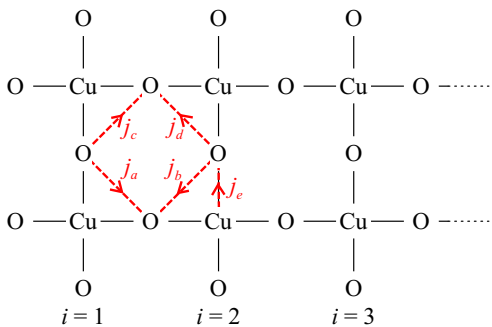}
\caption{
Schematic lattice structure of a two-leg CuO ladder.
$j_{r}\ (r=a,b,c,d,e)$ denote various local current operators
considered in this work. $i$ is the rung index. 
}
\label{lattice}
\end{figure}

In this paper we supplement our previous studies of the three-band model for 
two-leg CuO ladders~\cite{Nishimoto02,Jeckelmann98} 
by an analysis of the current-current correlation 
functions in doped systems for various parameters $\Delta_{pd}$ and $V_{pd}$. 
The hole Hamiltonian for this model is given by
\begin{eqnarray}
H &=& 
- t_{pd} \sum_{\langle
ij\rangle\sigma} \left(d^\dagger_{i\sigma}
p_{j\sigma}^{\phantom{\dagger}} + p^\dagger_{j\sigma}
d_{i\sigma}^{\phantom{\dagger}} \right) \nonumber \\
&& - t_{pp} \sum_{\langle ij\rangle\sigma}
\left(p^\dagger_{i\sigma}p_{j\sigma}^{\phantom{\dagger}} +
p^\dagger_{j\sigma}
p_{i\sigma}^{\phantom{\dagger}} \right) \label{hamiltonian} \\
&& +U_d \sum_i d^\dagger_{i\uparrow}
d_{i\uparrow}^{\phantom{\dagger}} d^\dagger_{i\downarrow}
d_{i\downarrow}^{\phantom{\dagger}}
+ U_p \sum_i p^\dagger_{i\uparrow}
p_{i\uparrow}^{\phantom{\dagger}} p^\dagger_{i\downarrow}
p_{i\downarrow}^{\phantom{\dagger}} \nonumber \\
&& + V_{pd} \sum_{\langle
ij\rangle\sigma\sigma^\prime} p^\dagger_{i\sigma}
p_{i\sigma}^{\phantom{\dagger}} d^\dagger_{j\sigma^\prime}
d_{j\sigma^\prime}^{\phantom{\dagger}} + \Delta_{pd} \sum_{i\sigma}
p^\dagger_{i\sigma} p_{i\sigma}^{\phantom{\dagger}} ,  \nonumber
\end{eqnarray}
where the operators 
$d_{i\sigma}^{\dagger}$ and $p_{i\sigma}^{\dagger}$ create holes 
with spin $\sigma$ in the Cu $d$-orbitals and  the O $p$-orbitals, respectively.  
The geometry of the two-leg CuO ladder is illustrated 
in Fig.~\ref{lattice}, where the rung and leg O sites represent $p_y$ and  $p_x$ 
orbitals, respectively.
The first and fifth sums are over all nearest-neighbor Cu-O pairs while
the second sum is over all nearest-neighbor $p_x$-$p_y$ pairs on O sites. 
The index $i$ runs over all Cu sites in the third term and over all O sites
in the fourth and sixth sums.
$t_{pd}$ is the hopping integral between nearest-neighbor Cu and O sites 
(solid lines in Fig.~\ref{lattice}) 
and $t_{pp}$ is the hopping integral between nearest-neighbor $p_x$-$p_y$ pairs on O sites. 
We have chosen the phases of the orbitals such that the signs of the 
hopping matrix elements are constant. With the minus sign convention 
of Eq.~(\ref{hamiltonian}) one has
$t_{pd}>0$ and $t_{pp}\geq 0$. $U_d$ and $U_p$ 
are the on-site Coulomb energies for Cu and O sites, respectively.
% and the energy difference between the O and Cu sites is
%$\Delta_{pd}=\epsilon_p-\epsilon_d \ge 0$. We also consider the 
%Coulomb interaction $V_{pd}$ between nearest-neighbor Cu and O orbitals.
We will work in units where $t_{pd}=1$
and use the typical values $t_{pp}=0.5$, $U_d=8$, and $U_p=3$,
throughout.~\cite{Hybertsen89} 
In this model an undoped CuO-ladder corresponds to a density of one hole per 
Cu site. The hole concentration per Cu atom is $x=1+y$ with the doping
rate $y=N/(2L)$, where $N$ is the number of doped holes ($N>0, y >0$) or 
doped electrons ($N<0, y < 0$) in a ladder with $L \times 2$ Cu atoms. 

In this work correlation functions are calculated numerically using the density-matrix
renormalization group (DMRG)
method.~\cite{White92, Schollwoeck05} We have used up to $m=3400$ density-matrix 
eigenstates to build the DMRG basis. The DMRG ground-state energy is 
estimated to be accurate to parts in $10^{-2}t_{pd}$ or better for
ladders with open boundary conditions and up to $40 \times 2$ Cu sites
[corresponding to a total of $282$ sites (Cu or O)].
As in Ref.~\onlinecite{Greiter07}
we have calculated correlations 
\begin{equation}
%c_{rs}(i_1,i_2)=\left\langle j_{r}^{\phantom{\dagger}}(i_1)j_{s}^\dagger(i_2) \right\rangle, \ 
c_{rs}(i_1,i_2)=\left\langle j_{r}(i_1)j_{s}(i_2) \right\rangle, \ 
\end{equation}
of various local currents $j_{r}$ between nearest-neighbor O sites or nearest-neighbor Cu-O pairs
which are illustrated in Fig.~\ref{lattice}.
%The operators $j_{a,b}(i)$ are proportional to the current operators and defined by 
%\begin{eqnarray}
%j_{a}(i)& = &\sum_\sigma p_{xi\sigma}^{\dagger} p_{yi\sigma}^{\phantom{\dagger}}
%-p_{yi\sigma}^{\dagger} p_{xi\sigma}^{\phantom{\dagger}}, \nonumber\\ 
%j_{b}(i)& =& \sum_\sigma p_{yi+1\sigma}^{\dagger} p_{xi\sigma}^{\phantom{\dagger}}
%-p_{xi\sigma}^{\dagger} p_{yi+1\sigma}^{\phantom{\dagger}},
%\label{current}
%\end{eqnarray}
%where $p_{yi\sigma}$ annihilates a hole with spin $\sigma$ 
%in the $p_y$-orbital of the O site in the middle of the rung with index $i$
%and $p_{xi\sigma}$ acts similarly on the $p_x$-orbital of the O site 
%between the rungs $i$ and $i+1$ in the lower leg. 
We have not found any long-range ordered current patterns 
for any set of the model parameters that we have investigated. 
Current-current correlations always decay faster than $1/l$ as a function of 
distance $l=|i_2 - i_1|$ in the present two-leg ladder system. 
Despite the absence of long-range order we can search for patterns in the sign of the
correlation functions $c_{rs}$. 
Close to the ladder ends these signs fluctuate as widely as
in small square clusters.~\cite{Greiter07} 
In the middle of long enough ladders, however, the sign of a given function $c_{rs}$
does note change with the distance $l=|i_2 - i_1|$. 
In that case, the relative phases of the current-current 
correlations for various directions are compatible with the translationally
invariant CC-pattern $\theta_I$ proposed by Varma.~\cite{Varma06}
We conclude that two-leg CuO ladders have CC-like current fluctuations. 

The various correlation functions $c_{rs}$ show a qualitatively similar 
dependence on the interaction parameters and the hole concentration. 
Therefore, we will discuss only $c_1 = c_{aa}$ hereafter. 
As open boundary conditions are used, 
the correlation functions $c_1(l) = c_{aa}(i_1,i_2)$ 
have been calculated using distances $l=|i_1-i_2|$
taken about the midpoint of the ladders 
(i.e., the integer part of $\frac{i_1+i_2}{2}$ equals $L/2$).
We only show results for $l \lesssim L/2 = 20$ which have been
obtained in ladders with $40 \times 2$ Cu atoms, so that edge effects are small. 

\begin{figure}[t]
    \includegraphics[width=5.5cm,clip]{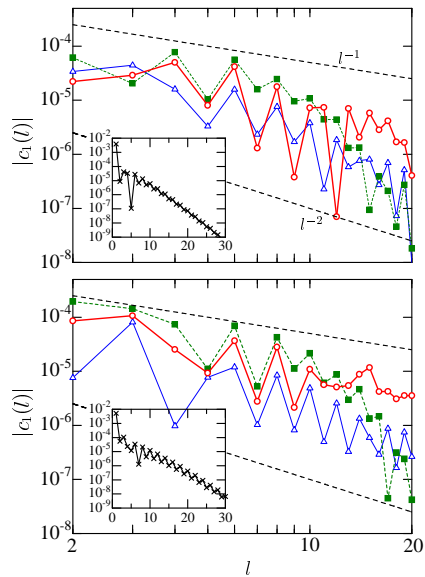}
  \caption{
Log-log plot of the correlation function $|c_1(l)|$ versus $l$ for an electron-doped 
(top panel) and a hole-doped (bottom panel) ladder 
with $\Delta_{pd}=3$ and $V_{pd}=0$. 
Triangles, squares, and circles correspond to four, six, 
and eight doped particles ($|y|=5, 7.5, 10$ \%), respectively.  
Lines are guides for the eyes.  The dashed lines have slope $-1$ and $-2$. 
Inset: Semilog plot of $|c_1(l)|$ for a ladder doped with two particles ($|y|=2.5$\%).
  }
    \label{D3V0_fig}
\end{figure}

We first investigate the evolution of the current-current correlations 
upon doping. Some results for 
$c_1(l)$ versus $l$ are shown in Fig.~\ref{D3V0_fig} for various 
hole concentrations $x$. 
Although there are substantial differences between hole-doped and electron-doped
two-leg ladders in the three-band model,~\cite{Nishimoto02} 
we have found that the current-current 
correlations are qualitatively similar in both cases.
In systems doped with two holes or two electrons
these correlations 
decay exponentially with distance (see the inset of Fig.~\ref{D3V0_fig}). 
This behavior can be seen for all interaction parameters that we have used.  
In ladders doped with at least four electrons or holes ($|y| \geq 5$  \%), however, 
we have found that current-current correlation functions  
exhibit an approximate power-law decay 
$l^{-\nu}$ with $1 < \nu \le 2$ for $l \gtrsim 3$. The overall magnitude
of the correlation function $|c_1(l)|$ is larger for six or eight doped particles
than for four doped particles
We therefore conclude that current-current fluctuations 
are enhanced upon doping.

\begin{figure}[t]
    \includegraphics[width= 5.5cm,clip]{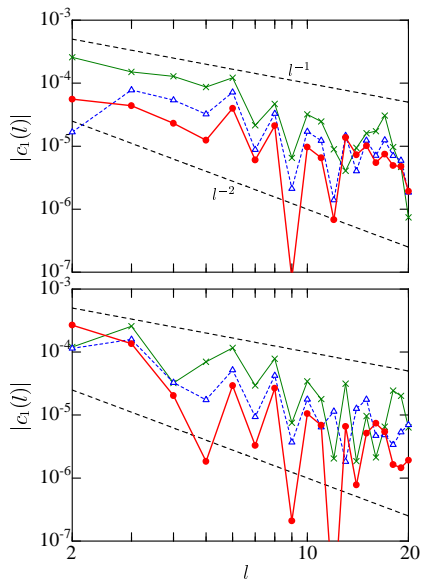}
  \caption{
Correlation function $|c_1(l)|$ versus $l$ for an electron-doped (top panel) 
and a hole-doped (bottom panel) ladder ($|y|=10$\%) with $\Delta_{pd}=2$. Crosses, 
triangles, and circles correspond to $V_{pd}=0$, $1$, and $2$, respectively.
Lines are guides for the eyes.  The dashed lines have slope $-1$ and $-2$. 
  }
    \label{D2_fig}
\end{figure}

We next turn to the effect of the Coulomb interaction $V_{pd}$ between nearest-neighbor
Cu and O sites.
Figure~\ref{D2_fig} shows the current-current 
correlation function $|c_1(l)|$ versus $l$ for $V_{pd}=0,1,2$. 
(A recent {\it ab initio} calculation~\cite{Korshunov05} suggests
that $V_{pd} \sim 1-1.5$ is appropriate for cuprates.)
The doping is $|y| = 10$ \% (eight doped electrons or holes) 
which is close to optimal doping in high-temperature superconducting cuprates.
We see that the results are similar for hole and electron doping and do not
substantially change as a function of $V_{pd}$. 
Although the overall amplitude of $|c_1(l)|$ is slightly reduced by increasing $V_{pd}$, 
its order of magnitude does not change from $V_{pd}=0$ to $2$. 

\begin{figure}[t]
    \includegraphics[width=5.5cm,clip]{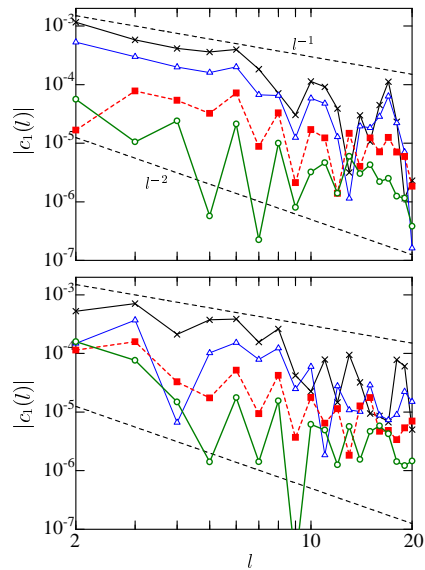}
  \caption{
Correlation function $|c_1(l)|$ versus $l$ for an electron-doped (top) 
and a hole-doped (bottom) ladder ($|y|=10$ \%) with $V_{pd}=1$. 
Crosses, triangles, squares, and circles correspond to $\Delta_{pd}=0$, $1$, $2$, and $3$, 
respectively.
Lines are guides for the eyes.  The dashed lines have slope $-1$ and $-2$. 
  }
    \label{V1_fig}
\end{figure}

The current-current correlation depends more significantly on the energy 
difference $\Delta_{pd}$ than on the Coulomb repulsion $V_{pd}$. 
In Fig.~\ref{V1_fig}, we show $|c_1(l)|$ versus $l$ for several values of $\Delta_{pd}$. 
While it has been generally accepted~\cite{Hybertsen89} that 
$\Delta_{pd}=2-3$, Varma has proposed~\cite{Varma06} that the CC 
patterns are stabilized only when $\Delta_{pd} \lesssim {\cal O}(t_{pd})$.
We have indeed found that the amplitude of the current-current correlations decreases with 
increasing $\Delta_{pd}$. For $\Delta_{pd}=3$, $|c_1(l)|$ is 
an order of magnitude smaller than for $\Delta_{pd}=0$. 

\begin{figure}[t]
    \includegraphics[width= 6.5cm,clip]{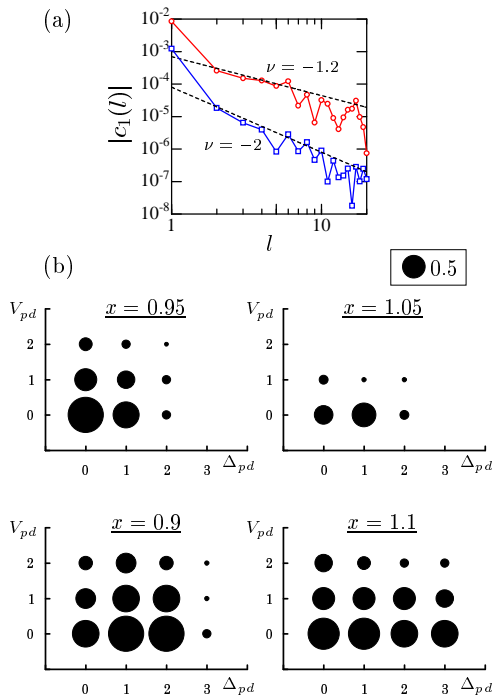}
  \caption{
(a) Correlation function $|c_1(l)|$ for $(\Delta_{pd}, V_{pd}, x)=(2, 0, 0.9)$ (circle)
and $(3, 2, 0.95)$ (square). The dashed lines are the fitted functions $A l^{-\nu}$. 
(b) Fitted values of $\nu$ in the parameter space $(\Delta_{pd}, V_{pd})$ 
for various hole concentrations $x$.  
The radius of the solid circles is proportional to $2-\nu$. 
  }
    \label{fig5}
\end{figure}

Our data show that the overall amplitude of current fluctuations (for $l \leq 20$)
increases with doping, decreases markedly with increasing $\Delta_{pd} \leq 3$ 
but is little affected by $V_{pd}$.
However, to understand the long-range behavior of power-law correlations, 
it is necessary to investigate the variations of their exponent $\nu$.
We have estimated $\nu$ by fitting our numerical data for the correlation function
$|c_1(l \geq 2)|$ to a function $A l^{-\nu}$, where both $A$ and $\nu$ are fit parameters.
As an illustration Fig.~\ref{fig5}(a) 
shows two such fits: The first one corresponds 
to a rapid decay ($\nu\approx 2$) of the current-current correlations while
the second one yields one of the smallest exponent, $\nu \approx 1.2$, that we have found.
As we use data for short distances $l \leq 20$ only and the correlation functions
oscillate widely, the fitted values of $\nu$ are not quantitatively accurate.
Nevertheless, we think that the variations of the fitted exponent $\nu$
give a qualitative indication of the variations in
the long-range behavior of the corresponding correlation functions. 

Figure~\ref{fig5}(b) shows the fitted values of $2-\nu$ for the current-current 
correlations $c_1(l)$ in the parameter space 
($\Delta_{pd}$, $V_{pd}$) for several hole concentrations. 
(We use the deviation of the exponent from its value in a Fermi sea, $2-\nu$,
as a measure of the strength of current-current correlations.) 
Two clear trends can be observed both for electron and hole dopings: 
The current fluctuations
decrease faster for low doping $|y|=5$ \% than for high doping $|y|=10$ \%
and $\nu$ increases with the nearest-neighbor coupling $V_{pd}$. 
The dependence of $\nu$ on the energy difference $\Delta_{pd}$ is irregular
but a large value $\Delta_{pd} \geq 3$ results in a rapid decay of 
current-current  correlation functions.
The smallest exponent $\nu \approx 1.2$ is found around $\Delta_{pd}=1-2$ for electron doping.
For hole doping, however, the smallest exponent $\nu \approx 1.3$ 
is found for $\Delta_{pd}=0-1$.

In our previous work~\cite{Nishimoto02}
we studied the pairing correlations in the three-band model
for two-leg CuO ladders. We found that electron- and hole-doped systems exhibit
$d$-wave-like power-law pairing correlations. 
Therefore, 
the three-band model at low-doping ($|y| \alt 2.5$\%) and the $t-J$ model 
have similar properties: 
Power-law pairing correlations and exponentially decaying current 
fluctuations.~\cite{Scalapino01}
At high enough doping, however, 
both pairing and current power-law fluctuations
seem to coexist in the three band model.  
Comparing fitted exponents for the current and pairing correlation
functions in hole doped ladders ($ 5 \leq y \leq 10$ \%),
we find that pairing correlations always
dominate (i.e., decay significantly slower) 
for $\Delta_{pd} \geq 2$ while current correlations dominate only
in a small region of parameter space ($\Delta_{pd} \leq 2, V_{pd} \leq 1$) 	
at the highest doping rate investigated ($y = 10$ \%). 
Thus in the hole-doped three-band model on two-leg ladders, there
is a region of enhanced and apparently dominant current fluctuations 
in good agreement with the interaction parameters proposed by 
Varma.~\cite{Varma06}

For these parameters, however, we showed in our previous 
study~\cite{Nishimoto02} that
the undoped ladder has only very small charge and spin 
gaps (which probably vanish in the limit of infinitely long ladders). Moreover,
local spin moments are not formed
on the Cu sites as holes are not localized on those sites at any doping, and thus
there is no tendency toward (short-range) antiferromagnetic order between the Cu sites.
Therefore, in the regime of the three-band model, where dominating current fluctuations 
are found in two-leg ladders, the undoped system is  
a paramagnetic metal or small-gap insulator.
In the regime of the three-band model where undoped ladders are
``antiferromagnetic'' insulators (see Ref.~\onlinecite{Nishimoto02})
current fluctuations are not enhanced
($\nu \approx 2$) or decay faster than pairing correlations.
We conclude that no phase with CC order or dominating CC fluctuations
occurs in the three-band model on a two-leg ladders with realistic
parameters for cuprate compounds.

\acknowledgments

We thank T. Giamarchi, R. Thomale, and P. W\"{o}lfle for helpful discussions.

\end{document}